\documentclass[12pt]{article}

\usepackage{epsfig}

\title{\small \bf Leptogenesis and tensor polarisation from a gravitational Chern-Simons term}

\author{\small David H. Lyth$^{1,} \ $\footnote{Email address: d.lyth@lancaster.ac.uk} \ ,
Carlos Quimbay$^{2,} \ $\footnote{Associate researcher of the Centro Internacional
de F\'{\i}sica, Ciudad Universitaria, Bogot\'a D.C., Colombia. Email address:
cjquimbayh@unal.edu.co} \ , and Yeinzon Rodr\'{\i}guez$^{1, \ 3,} \ $\footnote{Email
address: y.rodriguezgarcia@lancaster.ac.uk} \\
\small $^1$ Department of Physics, Lancaster University,\\
\small Lancaster LA1 4YB, UK\\
\small $^2$ Departamento de F\'{\i}sica, Universidad Nacional de Colombia,\\
\small Ciudad Universitaria, Bogot\'a D.C., Colombia \\
\small $^3$ Centro de Investigaciones, Universidad Antonio Nari\~no,\\
\small Cll 58A \# 37-94, Bogot\'a D.C., Colombia}

\date{\small January 2005}

\usepackage[activeacute,english]{babel}
\usepackage{latexsym}

\newcommand{\bn}{\begin{eqnarray}}
\newcommand{\en}{\end{eqnarray}}
\newcommand{\be}{\begin{equation}}
\newcommand{\ee}{\end{equation}}

\newcommand{\la}{\label}

\newcommand{\ci}{\cite}
\newcommand{\bea}{\begin{eqnarray}}
\newcommand{\eea}{\end{eqnarray}}

\renewcommand\({\left(}
\renewcommand\){\right)}
\renewcommand\[{\left[}
\renewcommand\]{\right]}

\newcommand\eq[1]{Eq.~(\ref{#1})}

\newcommand\GeV{\,\mbox{GeV}}

\newcommand\mpl{M_{\rm P}}

\newcommand{\lsim}{\mbox{\raisebox{-.9ex}{~$\stackrel{\mbox{$<$}}{\sim}$~}}}
\newcommand{\gsim}{\mbox{\raisebox{-.9ex}{~$\stackrel{\mbox{$>$}}{\sim}$~}}}

\def\caln{{\cal N}}

\newcommand\bfk{{\bf k}}

\newcommand\bfx{{\bf x}}

\newcommand\sub[1]{_{\rm #1}}

\begin{document}

\maketitle

\begin{abstract}

\noindent 
Within an effective field theory derived from string theory,
the universal axion has to be coupled to the the gravitational Chern-Simons (gCS) term.
During any era when the axion field is varying, the vacuum 
fluctuation of the gravitational wave amplitude will then be circularly polarised,
generating an expectation value for the gCS term. The polarisation may be observable
through the Cosmic Microwave Background, and the vacuum expectation value of the gCS
term may generate the baryon asymmetry of the Universe. We argue here that such effects
cannot be computed without further input from string theory, since the `vacuum' in question
is unlikely to be the field-theoretic one. 

\end{abstract}

\subsection*{1. Introduction}
It has been pointed out \cite{LWK,CHH,SPS,SJG,SJM,polchinski} 
that the action of the effective field theory
should contain a gravitational Chern-Simons (gCS) term
coupled to a scalar field such as the
universal axion:
\bea
S\sub{gCS} & =&  \frac{\mpl^2}{4} \int d^4x 
f(\phi) R \tilde R \,, \\
R \tilde R & \equiv &
{\epsilon}^{\alpha \beta \gamma \delta} R_{\alpha \beta \rho \sigma} 
{R_{\gamma \delta}}^{\rho \sigma} \,,
\eea
as a result of the Green-Schwarz mechanism \cite{GS}.
Here $\mpl =2.4\times 10^{18}\GeV$ is the reduced Planck mass,
and  $f$ is an odd function.

In the reasonable approximation that
$f$ is linear it may be written as
\be
f = \frac \caln {\mu^2} \frac\phi\mpl \,, \label{fexp}
\ee
where  $\mu$  is the string scale (representing 
the ultra-violet cutoff for the effective field theory including gravity)
and $\caln$ is of order 1 at least for the case studied in \cite{SJG}\footnote
{Our  $\caln$ is $(\mu/\mpl)^2$ times that of Ref. \cite{SJM}.}.

If $f$ varies during some era, the gCS term will polarise the 
vacuum fluctuation of the gravitational wave amplitude.
On cosmological scales, this may be directly
observable through the Cosmic Microwave Background 
\cite{LWK,CHH,SJM}. On much smaller scales
it may be observable indirectly, through the generation of baryon number 
\cite{SPS,SJG}. In this note we examine both effects.

\subsection*{2. Quantising the tensor perturbation}
Quantisation of the gravitational wave amplitude in the presence of a
gCS term has not been given before. We describe it here, adopting an
approach which relies mainly on the field equation.

We write the line element for the expanding Universe, displaying only the 
tensor perturbation which corresponds to a gravitational wave
amplitude:
\be
ds^2 = a^2(\tau) \left( d\tau^2 - \left(\delta_{ij}+ 2h_{ij}
(\bfx,\tau) \right)dx^idx^j \right).
\ee

Working to first order, the
transverse and traceless gravitational wave amplitude $h_{ij}$
may be written as
\be 
h_{ij}  ({\bf x}, \tau) =
\frac{\sqrt{2}}{M_{Pl}}
\int  \frac{d^3 k}{(2\pi)^{3/2}} e^{i {\bfk \cdot \, \bfx}} \sum_p 
\epsilon_{ij}^p(\bfk) h_p({\bf k},\tau)
\label{class} \,,
\ee
where $p=$ R or L, and the polarisation tensors satisfy\footnote
{These expressions follow from the
properties of the rotational transformations.}
\bea
k_i \epsilon_{ij}(p,{\bf k}) &=& 0, \nonumber \\
\epsilon^\ast_{ij}(p,{\bf k})\epsilon_{ij} (p',{\bf k}) &=&
2\delta_{pp'}, \nonumber \\
\epsilon^{ilm} \epsilon_{ij}^{\ast}(L, {\bf
k}) \epsilon_{jl}(R, {\bf k}) &=& \epsilon^{ilm}
\epsilon_{ij}^{\ast}(R, {\bf k}) \epsilon_{jl}(L, {\bf
k}) = 0, \nonumber \\
\epsilon^{ilm} \epsilon_{ij}^{\ast}(L, {\bf k}) \epsilon_{jl}(L,
{\bf k}) &=& -\epsilon^{ilm} \epsilon_{ij}^{\ast}(R, {\bf k})
\epsilon_{jl}(R, {\bf k}) = -2i\frac{k_m}{\left| {\bf k} \right|}. 
\label{nprop2}
\eea

We first recall the familiar situation where there is no gCS term,
presenting it in a way which will make it easy to include the gCS term.
We need the  field equation for the mode functions $h_p$, evaluated
to first order so that it is linear in $h_p$. It is given by
the  Einstein action
\be
S\sub E = \frac{\mpl^2} 2 \int d^4x (-g)^{1/2} R \,,
\ee
and it may be obtained by either of two routes. 
\begin{enumerate}
\item  Start with  the full field equation $R_{\alpha\beta}=0$ and take its
first-order perturbation \cite{lifshitz}.
\item Take the second-order perturbation of the full action
and  from it derive the first-order field equation \cite{starob}. 
\end{enumerate}
Either way, it is convenient to consider a re-defined mode function
\be
\mu_p \equiv z_p h_p
\label{mudef} \,,
\ee
with\footnote{We use the notation $z_p$ because this quantity will 
depend on the polarisation on $p$ when we include the gCS term.} $z_p\equiv a$.
Then the field equation  is 
\be
\mu_p'' + \( k^2 -\frac{z_p''}{z_p} \) \mu_p = 0
\label{mueq} \,,
\ee
where a prime will denote $d/d\tau$.

To quantise one needs an action.
The second-order action  is given by
$S\sub E=\int d\tau L\sub E$, with lagrangian
\be
L\sub E = \frac12 \int d^3k  \[ \mu_p '{}^2 +
\(  k^2  + \frac{z_p''}{z_p} \) \mu_p^2 \]
\label{leq} \,.
\ee
This action is determined (up to irrelevant surface terms)
by the field equation except for its normalisation. To determine that 
normalisation it is enough to know the action in the subhorizon regime
$k\gg aH$. In this regime  the term $z_p''/z_p$ becomes negligible
and,  the action as well as the field equation
is  the same as for a massless scalar
field with Fourier components $\mu_p$. We see that the first approach,
supplemented by an evaluation of the second-order action in the sub-horizon
regime,  can provide the full second-order action\footnote
{It is even superfluous to write down the action except in the sub-horizon
regime since canonical normalisation once established is preserved by the
equation of motion. Similar considerations apply to the case of the curvature perturbation
\cite{book}. In both cases, the second approach has become
the standard one, though the simpler first approach was the original one.}.

Promoting  the gravitational wave amplitude to an operator $\hat h_{ij}$,
we work in the Heisenberg picture so that the operator satisfies the 
classical equation of motion.
Expanding in  Fourier modes we write
\be \hat{h}_{ij}({\bf x}, \tau) =
\frac{\sqrt{2}}{M_{Pl}} \int  \frac{d^3 k}{(2 \pi)^{3/2}} 
\sum_p \left[e^{i {\bf k \cdot \, x}}h_p( k,\tau)
\epsilon_{ij}(p,{\bf k})\hat{a}_p({\bf k}) \right] \la{Fou1}. \ee
The mode function $h_p(k,\tau)$ is related to the 
operator $\hat h_p(\bfk,\tau)$ appearing in the expansion of \eq{class} 
by
\be
\hat h_p(\bfk,\tau) = h_p (k,\tau) \hat a_p(\bfk) +
h_p^*(-k,\tau) \hat a_p^\dagger(-\bfk) \,.
\ee

Without loss of generality we impose the commutation relation
\be
[\hat a_p(\bfk),\hat 
a_{p'}^\dagger(\bfk') ] = \delta^3(\bfk-\bfk') \delta_{pp'} \,.
\label{com}
\ee
Then canonical quantisation corresponds to choosing the following
Wronskian for the mode function;
\be
\mu_p' \mu_p^* - \mu_p {\mu_p^*}' = -i
\label{wr} \,.
\ee
We also may define a time-independent vacuum state by
\be
\hat a_p(\bfk) |0 \rangle = 0
\label{vac} \,.
\ee
Finally, we 
demand that well before horizon entry
the vacuum corresponds to the state with no gravitinos,
which corresponds to choosing the mode function
\be
\mu_p = \frac1{\sqrt{2k}} e^{-ik\tau}
\label{initial} \,.
\ee
The mode function at later times is calculated by solving 
\eq{mueq}, but we shall not need it.

With these preliminaries it is easy to include the effect of the gCS
term. Without loss of generality we can continue to define the vacuum by
\eq{vac}, but now the mode function is different.
Either from the field equation $R_{\alpha\beta}
= \tilde T_{\alpha\beta}$, where the right hand side is the effective
energy-momentum tensor provided by the gCS term \cite{CHH}, or by deriving
the contribution of the gCS term to the second-order action \cite{SJM},
one can show that the mode functions $\mu_p$ defined by \eq{mudef}
satisfy the field equation in \eq{mueq}, with the modified factors
\be
z_p = \sqrt{a^2 \pm k f'} \,,
\ee
where the plus sign is for $p=$ L and the minus sign is for $p=$ R.

The total second-order action (coming from
$S\sub E + S\sub{gCS}$) has the  form \eq{leq}, up to irrelevant surface
terms. Indeed, by analogy with the previous discussion, the action
is defined up to a constant factor by the field equation, and the
factor is defined by matching to an already-known limit, namely the 
limit  $f'\to 0$ in which the gCS term is negligible. 






The above argument, leading to the second-order action in the presence
of a gCS term has not been given before. An expression for this action,
derived from a lengthy calculation, has been given in \cite{SJM},
but it is not manifestly equivalent (up to surface terms) to \eq{leq}.
On the basis of the above argument we know that it must be equivalent,
but there is no need to show this explicitly.

Since the action is the same as before, the commutation relation in
\eq{com} and the Wronskian condition in \eq{wr} are also the same and we can
still define the vacuum by \eq{vac}. The crucial difference, as we are now going to discuss,
is that there is unlikely to be  any initial sub-horizon regime where  the gCS term is 
negligible (unless it is negligible at all times in which case there is 
nothing to discuss). In the absence of such a regime, the initial condition for the mode functions
is at present unknown which means that one cannot calculate anything.

\subsection*{3. The generation of circularly polarised gravitational waves}
During inflation, comoving scales  comparable with the
present Hubble distance are 
supposed to leave the horizon\footnote{Recall that wavenumbers $k< aH$ are said
to be outside the horizon, and bigger wavenumbers are said to be inside the horizon.}.
The vacuum fluctuation of the gravitational wave amplitude on these scales is then converted to a classical perturbation, which after
horizon entry corresponds to  gravitational waves whose effect on the Cosmic
Microwave Background anisotropy may be observable. The idea  \cite{LWK,CHH,SJM} is that the
gCS term will give the  waves some circular polarisation whose effect may be detectable.

The amplitude of the gravitational waves is proportional to the 
inflationary Hubble parameter $H_*$, and the present bound 
$r\lsim 10^{-1}$
on their spectrum relative to that of the curvature perturbation
requires $H_* \lsim 10^{-5}\mpl$. On the other hand, they will never
be detectable \cite{CMB} unless $r\gsim 10^{-4}$, corresponding to
$H_*\gsim  10^{-6}\mpl$.   
We focus on this high range for $H_*$, while recognising that most inflation
models give a smaller value \cite{book} corresponding to unobservable gravitational
waves.

There are  a three  things to take into account, which have perhaps
not been sufficiently emphasised before.
\begin{enumerate}
\item The effective field theory will presumably become valid only 
when the energy density $3\mpl^2 H^2$ falls below $\mu^4$.
\item The  discussion for a given  scale $k$
should begin only when the wavenumber is below the 
string scale, $k/a < \mu$.
\item In order to obtain a prediction using known methods, we need 
an initial era when the gCS term has  negligible time-dependence, so that
the initial condition in \eq{initial} can be imposed on the mode function.
\end{enumerate}
The problem, as we now explain,  is that
the second and third requirements are practically incompatible if the 
polarisation is to have an observable effect.  

Taking $f$ to  have the linear form as in \eq{fexp}, $z_p$ is then given by
\be
(z_p/a)^2 - 1
=  \pm  \caln \( \frac{k}{a\mu} \) \(\frac{\mpl H}{\mu^2} \)
\( \frac \mu\mpl \) \( \frac{\dot\phi}{\mpl H} \)
\label{zeq2} \,.
\ee
The term on the right comes from the gCS term, and to satisfy our third
requirement we would need  an initial era when this term
 is negligible. 

The first, second, and third terms in brackets are obviously less than 1.
Provided that $\phi$ is canonically normalised the final term is also less
than 1, because the energy density then has a contribution 
$\frac12 \dot\phi^2 $ which must be less than the total energy density.
(For an estimate it is reasonable to suppose that one can take $\phi$ to be
approximately canonically normalised while it is varying.
Doing that  might be artificial from the string viewpoint,
altering the  expectation $\caln\sim 1$, but according to \cite{SJG}
one still expects $\caln \lsim 100$ or so.) If $\phi$ is the inflaton,
or a component of the inflaton in a multi-field model, the final term  
is very much less than 1. In addition it is slowly varying so that it may
be regarded as a constant when evaluating $z_p''$.

We see that  the effect of the gCS term will be significant only if the
smallness of the product of the four terms in brackets is compensated by
a sufficiently large value of $\caln$. Assuming that this happens,
we will now explain why the requirements 2 and 3 are likely to be incompatible.

The scale $k$ in which we are interested (corresponding to the size of
the presently observable Universe)
leaves the horizon during inflation at the 
 epoch  $k=aH_*$. 
This  scale emerges from the string scale
(at the epoch $k=a\mu$)  only
$N\sub{str} \equiv 
\ln (\mu/H_*)< \ln(\mpl/H_*)$ $e$-folds before it leaves the horizon, 
unless inflation begins a fewer number of $e$-folds before that epoch.
But to  get an observable  tensor perturbation one needs 
 $H_*\gsim  10^{-6}\mpl$,
making  $N\sub{str} < 14$. 
It seems unlikely that inflation will start
such a small number of $e$-folds before the observable Universe leaves
the horizon. Discounting that possibility, we see that {\em on the 
scale of interest, the gCS
term can be significant only during the first few Hubble times
after it emerges from the string scale}. This is why requirements
2 and 3 are practically incompatible; they require that the
motion of $\phi$ switches on suddenly during inflation, within just the
few Hubble times before the observable Universe leaves the horizon.

Having explained why the gCS term will not be initially negligible
(unless it is always so) let us consider the opposite possibility
that it initially dominates making $|z_p/a|^2 \gg 1$.
Then  evolution will   become  singular \cite{SJM}
when  $(z_p/a)^2$ passes through zero for one of the polarisation states.
At this point,  at least the linear calculation becomes unphysical.

It may be that a non-linear calculation would make sense, allowing
$|z_p/a|^2 \gg 1$ as an initial condition.
It has been noticed \cite{SJM} that 
 the initial value of 
$z_p''/z_p$ is then practically zero, 
if $\dot\phi$ is practically
constant corresponding to $\phi$ rolling very slowly.
In that case
\eq{initial} is a correctly-normalised 
solution of the  mode function equation \cite{SJM}, 
but it seems hard to justify the  use of this 
solution as an initial condition. 
The physical mode function $h_p$ decreases like $a^{-1/2}$, as opposed
to the $a^{-1}$ behaviour that would correspond to a redshifting gravitino
momentum, which is hardly surprising since
it is the gCS term rather than Einstein gravity which is 
dominating the dynamics of $h_p$. As a result it is hard to see how the 
vacuum state in \eq{vac} can be regarded as a no-particle state representing
minimal energy density, which is the usual justification for using such
a state.
However we can still say that $\hat{a}$ annihilates the vacuum, as in \eq{vac},
at the expense of compensating this effect with an adequate, but at the moment
unknown, definition of the mode function. 
We conclude that even if the gCS term dominates initially,
one still does not know the initial condition for the mode function.

\subsection*{4. Leptogenesis}
It is known  \ci{AGW} that  the gravitational anomaly
violates lepton number conservation through a term
\be 
\partial_\mu J^\mu_L  \supset  
\frac{1}{16\pi^2} R \tilde R. \la{an1}
\end{equation} 
If the gCS  term is nonzero during some era $t_1<t<t_2$,  lepton number
$n_ L\equiv J^0_\ell$ is  generated:
\be
n_L = \frac1{16\pi^2} \int^{t_2}_{t_1} R \tilde R \, dt.
\ee
Alexander, Peskin, and Sheikh-Jabbari \ci{SPS} have pointed out that
this   could be the origin of the observed baryon number, with 
lepton number generating baryon number at the electroweak transition.

The idea is to generate the gCS term from the vacuum fluctuation
that we have been discussing. 
Up to second order, the contribution of  $h_{ij}$ to $R\tilde R$ is\footnote{
This expression reproduces 
the Eq. (10) in Ref. \cite{SPS} for the case where
gravity waves move in the $z$ direction.}
\be 
R \tilde R =
-\frac{8}{a^4} \, \epsilon^{ijk} \left( \frac{\partial^2}{\partial_l
\partial \tau}h_{jm} \frac{\partial^2}{\partial_m \partial_i}h_{kl}
- \frac{\partial^2}{\partial_l \partial \tau}h_{jm}
\frac{\partial^2}{\partial_l\partial_i}h_{km} +
\frac{\partial^2}{\partial \tau^2}h_{jl}
\frac{\partial^2}{\partial_i
\partial \tau}h_{lk} \right) \la{rr2}. \ee
Inserting \eq{Fou1}  the vacuum expectation value, 
given here for the first time, is\footnote{This expression is complex, corresponding to the fact that the 
naively-defined  operator $\widehat{R \tilde R}$ is not hermitian.
Presumably we should take the real part, corresponding to the 
hermitian part of the naively-defined operator.}
\bea
\langle 0 \mid \widehat{R \tilde R} \mid 0 \rangle 
= \frac{16}{a^4 \mpl^2} \int \frac{d^3k}{(2\pi)^3} \big[k^2 h_L^\ast (k,\tau)
h_L' (k,\tau) - k^2 h_R^\ast (k,\tau) h_R' (k,\tau) \nonumber \\
- h_L'^\ast (k,\tau) h_L'' (k,\tau) + h_R'^\ast (k,\tau) h_R'' (k,\tau)\big] \,. \label{vacR}
\eea
The integration over $k$ is cut off at the value $k\simeq a\mu$, 
and  the presence of spacetime derivatives 
presumably  will mean that  values close to the cutoff 
will dominate the integral.
If that is the case we need to know each mode function just after
it emerges from the string scale. Its  subsequent evolution is
irrelevant, assuming that it is not so drastic as to prevent
the value $k\simeq a\mu$ from dominating the integral.

Although it is only of historical interest, we end with the following
extended comment. The earlier discussions \cite{SPS,SJG} ignored the problem
that the initial condition
is unknown, and also dropped  terms  containing more than two spacetime derivatives
from the evolution equation. 
The appropriately normalised solutions
after the gCS term switches on are then
\be
h_p(k,\tau) = 
\frac1a e^{-ik\tau}e^{\pm k \Theta (\tau - \tau_0)/2},
\label{Hh}
\ee
where 
\be
\Theta \equiv  \frac{4}{a^2} \left(f'' + aH f'\right) \,.
\ee
Inserting this into \eq{vacR} gives\footnote
{In an earlier work \cite{LRQ} we  assumed
that the $+$, $\times$ states decoupled (instead of the L and R states),
arriving at the incorrect result that  $R\tilde R=0$. We thank the authors
of \cite{SPS} for pointing out this mistake.}
\begin{equation}
\langle 0 \mid \widehat{R \tilde R} \mid 0 \rangle 
=  \frac{4 H^2 \Theta}{\pi^2 
M_{P}^2} \mu^4 \,. \label{csfinal}
\end{equation}

This expression differs from the one obtained in 
\cite{SPS}.
The derivations cannot be directly compared because
we  have used the Fourier decomposition whereas those of \cite{SPS} 
used a Green function method.
As one easily checks, we would recover the results of \cite{SPS} 
if we multiplied \eq{Hh} by a factor $ik/H$,
but this would not correspond to canonical normalisation nor would
it satisfy the requirement that 
the factors $k$ and $a$ should enter into 
physical quantities only
through the physical wavenumber $k/a$.

\subsection*{5. Conclusion}
The initial condition should presumably be provided by string 
theory.
Since the presence of a gCS term is indeed a
requirement in
string theory, it seems urgent to either find the initial condition or
to become convinced that the universal axion is frozen until 
the energy density falls  well below the string scale
rendering  the  gCS term negligible.

\bigskip

\subsection*{Acknowledgments} 
D.H.L. is  supported by PPARC grants PPA/G/O/2002/00469,  PPA/V/S/2003/00104,
PPA/G/S/2002/00098 and PPA/Y/S/2002/00272.
C.Q. thanks Lancaster University for its hospitality and PPARC for its financial
support during his visit to Lancaster University. Y.R. wants to acknowledge
Lancaster University and Universities UK for their partial financial help and the
Colombian agencies COLCIENCIAS and COLFUTURO for their postgraduate scholarships.

\renewcommand{\refname}{{\large References}}

\end{document}